\documentclass[traditabstract]{aa} 
\usepackage{txfonts}
\usepackage{graphicx}
\usepackage{natbib}

\newcommand{\css}{\mbox{\object{CSS\,21055}}}
\newcommand{\tc}{$T_\mathrm{ecl}$}

\newcommand{\msun}{$M_\odot$}

\newcommand{\oc}{$O\!-\!C$}

\begin{document}
\title{The eclipsing post-common envelope binary CSS21055: a white
  dwarf with a probable brown-dwarf companion}

\author{
Beuermann, K. \inst{1} \and 
Dreizler, S. \inst{1} \and 
Hessman, F.~V. \inst{1} \and 
Backhaus, U. \inst{2} \and
Boesch, A. \inst{1} \and 
Husser, T.-O. \inst{1} \and
Nortmann, L. \inst{1} \and \\
Schmelev, A. \inst{1} \and
Springer, R. \inst{1}
} 

\institute{
Institut f\"ur Astrophysik, Georg-August-Universit\"at, Friedrich-Hund-Platz 1, D-37077 G\"ottingen, Germany \and
Didaktik der Physik, Universit\"at Duisburg-Essen,Universit\"atsstr. 2, 45117 Essen, Germany 
}
\date{Received 10 July 2013; accepted 19 September 2013}

\authorrunning{K. Beuermann et al.} 
\titlerunning{The eclipsing post-common envelope binary CSS21055}

\abstract{We report photometric observations\thanks{The light curves
    presented in Fig.\,1 are available as ASCII tables at the CDS via
    anonymous ftp to cdsarc.u-strasbg.fr (130.79.128.5) or via
    http://cdsweb.u-strasbg.fr/cgi-bin/qcat?J/A+A/... } of the
  eclipsing close binary CSS21055 (SDSS\,J141126+200911) that strongly
  suggest that the companion to the carbon-oxygen white dwarf is a
  brown dwarf with a mass between 0.030 and 0.074\,\msun. The measured
  orbital period is 121.73\,min and the totality of the eclipse lasts
  125\,s. If confirmed, CSS21055 would be the first detached eclipsing
  WD+BD binary. Spectroscopy in the eclipse could provide information
  about the companion's evolutionary state and atmospheric structure.
}


\maketitle


\section{Introduction}

In their list of eclipsing white-dwarf main-sequence binaries drawn
from the Catalina Sky Survey (CSS), \citet{drakeetal10} included the
Sloan Digital Sky Survey (SDSS) source \css\ (SDSS\,J141126+200911) as
a possible eclipsing close binary, without providing further
information, however. A single SDSS spectrum of the $g\!=\!17.80$
source is available (plate number 2771, fiber 0024, MJD\,54527)
\footnote{http://skyserver.sdss3.org/dr9/en/tools/chart/navi.asp},
which shows a white dwarf (WD) with a hydrogen spectrum of spectral
type DA. No signature of a late-type companion is seen. The observed
SDSS photometric AB magnitudes ($u\!=\!18.240(15)$,
$g\!=\!17.800(6)$, $r\!=\!17.981(7)$, $i\!=\!18.190(8)$, and
$z\!=\!18.428(27)$; Adelman et al. 2012) are
consistent with the spectral energy distribution of the WD and
provide no evidence of an infrared flux excess.
\citet{kleinmanetal13} derived an effective temperature
$T_\mathrm{1}\!=\!13\,074$\,K, gravity log\,$g\!=\!7.89$, and mass
$M_1\!=\!0.551$\,\msun, suggesting that it is a carbon-oxygen WD.

The binary nature was re-discovered by one of us (UB) using the
MONET/North telescope within the {\it PlanetFinders} project conducted
by high-school teachers and their students in collaboration with
professional astronomers. The light curves show narrow eclipses
suggestive of a brown dwarf (BD) companion.  If confirmed, this would
make CSS21055 one of the few post-common envelope binaries (PCEB), in
which a companion near or below the limit of stable hydrogen burning
has survived the common-envelope (CE) event that led to the birth of
the WD. The frequency of detached PCEB with WD and BD components is
small \citep{steeleetal11} and only very few are known so far. Of
these, only two have short orbital periods comparable with \css. These are 
NLTT5306 \citep{steeleetal13} and WD0137-349
\citep{maxtedetal06,burleighetal06}, which are not eclipsing. 

In this paper, we present time-resolved photometry of \css\ acquired
in 2012 and 2013 that allowed us to obtain an accurate ephemeris and
to derive the system parameters from a light-curve analysis.

\section{Observations and analysis}

\subsection{Observations}
\label{sec:obs}

The binary \css\ was observed with the 1.2-m MONET/North telescope at
the University of Texas McDonald Observatory via the MONET
browser-based remote-observing interface on 24 nights between March
2012 and March 2013, using an Apogee ALTA E47+ 1k$\times$1k CCD
camera. In June 2013, it was observed with the 1.9-m telescope
of the South African Astronomical Observatory (SAAO). The source is
located at $RA(2000)\!=\!14\,11\,26.2$, $DEC(2000)\!=\!+20\,09\,11.1$,
only $1\fdg4$ from Arcturus.  Photometry was performed relative to two
14-mag comparison stars C1 and C2 (SDSS\,J141125+201013 and
SDSS\,J141123.90+200813.2), located at 62\arcsec\,N, 8\arcsec\,W and
58\arcsec\,S, 42\arcsec\,W of CSS21055, respectively. We obtained
6.0\,h of data in white light (WL) with exposure times of 10\,s or
20\,s and 13.8\,h in the Bessell I-band (central wavelength 8000\,\AA)
with exposure times between 30\,s and 180\,s, all separated by 3\,s
readout. Our extensive observations provided full orbital coverage in
the \mbox{I-band} and about 50\% coverage in WL. Table~\ref{tab:monet}
provides information on those runs that cover the narrow eclipses,
with the last column indicating the telescope used. All additional
observations that contribute to the light curves outside the narrow
eclipse were performed with MONET/N.

\subsection{Ephemeris}
\label{sec:ephem}

The mid-eclipse times were determined from the individual light curves
using a uniform-disk model for the white dwarf and the secondary star
\citep[see][]{backhausetal12, beuermannetal13}. The formal
\mbox{1-$\sigma$} errors of the mid-eclipse times were calculated from
the measurement errors of the relative fluxes in the individual CCD
images, employing standard error propagation. We measured the
mid-eclipse times $t_\mathrm{ecl}$ using a fixed inclination of
$85\degr$ (see below). Table~\ref{tab:monet} lists the cycle numbers,
the mid-eclipse times, and their 1-$\sigma$ errors. The errors of \tc\
in WL vary between 0.3 and 1.6\,s, depending on the quality of the
individual light curves. The mean timing error is 0.70\,s.  All
mid-eclipse times were converted from UTC to Barycentric Dynamical
Time (TDB) and corrected for the light travel time to the solar system
barycenter, using the tool provided by
\citet{eastmanetal10}\footnote{http://astroutils.astronomy.ohio-state.edu/time/}.
The corrected times are quoted as Barycentric Julian Days in TDB in
Col.~4 of Table~\ref{tab:monet}. The linear ephemeris obtained from
the 21 WL mid-eclipse times is
\begin{equation}
T_\mathrm{ecl}\!=\!\mathrm{BJD(TDB)}~2455991.888717(2) + 0.0845327526(13)\,E.~~~
\label{eq:ephem}
\end{equation}
With $\,\chi^2\!=\!17.51$ for 19 degrees of freedom, there is no
evidence for a period change. The \oc\ values of the mid-eclipse times
relative to this ephemeris are quoted in Col.~5 of Table~\ref{tab:monet}.

\begin{table}[t]
\begin{flushleft}
  \caption{Mid-eclipse times of \css.}
\begin{tabular}{@{\hspace{0mm}}r@{\hspace{4mm}}c@{\hspace{4mm}}c@{\hspace{4mm}}c@{\hspace{4mm}}c@{\hspace{-0mm}}c}
\hline\hline \\[-1ex]
Cycle   &  JD         & Error & BJD(TDB)     & \oc\ & Tel\tablefootmark{1}\\
        & 2450000+   &  (days) & 2450000+    & (days) & \\[0.5ex]
\hline\\[-1ex]
\multicolumn{5}{l}{\emph{(a)~~~White light, (10\,s or 20\,s exposures)}}\\[0.5ex]
    0 & 55991.884052 & 0.000006 & 55991.888711 & \hspace{-2mm}$-$0.000006 & M\\
   25 & 55993.997279 & 0.000010 & 55994.002047 &  0.000011   & M\\
  261 & 56013.946247 & 0.000005 & 56013.951773 &  0.000007 & M\\
  273 & 56014.960600 & 0.000008 & 56014.966150 & \hspace{-2mm}$-$0.000009& M\\
  295 & 56016.820296 & 0.000007 & 56016.825886 &  0.000007  & M\\  
  296 & 56016.904811 & 0.000005 & 56016.910403 & \hspace{-2mm}$-$0.000009& M\\  
  297 & 56016.989349 & 0.000006 & 56016.994943 & \hspace{-2mm}$-$0.000002& M\\  
  307 & 56017.834658 & 0.000007 & 56017.840267 & \hspace{-2mm}$-$0.000005& M\\  
  308 & 56017.919200 & 0.000005 & 56017.924812 &  0.000007 & M\\
  344 & 56020.962325 & 0.000011 & 56020.967987 &  0.000003 & M\\
  356 & 56021.976700 & 0.000010 & 56021.982376 & \hspace{-2mm}$-$0.000002& M\\  
  521 & 56035.924560 & 0.000008 & 56035.930274 & \hspace{-2mm}$-$0.000007& M\\  
  650 & 56046.829451 & 0.000006 & 56046.835001 & \hspace{-2mm}$-$0.000006& M\\  
  651 & 56046.913994 & 0.000007 & 56046.919542 &  0.000003 & M\\
 1324 & 56103.807684 & 0.000009 & 56103.810081 & \hspace{-2mm}$-$0.000003 & M\\
 1335 & 56104.737627 & 0.000005 & 56104.739948 &  0.000006 & M\\
 1336 & 56104.822152 & 0.000011 & 56104.824466 & \hspace{-2mm}$-$0.000008& M\\  
 2244 & 56181.583545 & 0.000013 & 56181.580223 &  0.000009 & M\\
 4461 & 56368.984116 & 0.000009 & 56368.989324 & \hspace{-2mm}$-$0.000002& M\\
 4532 & 56374.985736 & 0.000008 & 56374.991149 & \hspace{-2mm}$-$0.000002& M\\
 5589 & 56464.339510 & 0.000019 & 56464.342276 &  0.000004 & S\\[1.3ex]
\multicolumn{5}{l}{\emph{(b)~~~Bessell I-band, (30\,s or 60\,s exposures)}}\\[0.5ex]
  119 & 56001.942981 & 0.000018 & 56001.948114 & \hspace{-2mm}$-$0.000000& M\\
  236 & 56011.832942 & 0.000015 & 56011.838413 & \hspace{-2mm}$-$0.000034& M\\  
  237 & 56011.917518 & 0.000020 & 56011.922991 &  0.000012  & M\\  
  355 & 56021.892136 & 0.000032 & 56021.897810 & \hspace{-2mm}$-$0.000034& M\\
 1240 & 56096.706399 & 0.000031 & 56096.709342 &  0.000012  & M\\ [1.3ex]  
 \hline\\[-1ex]
\end{tabular}

\tablefoottext{1}{\,M = MONET/North, S = SAAO 1.9-m telescope.}
\label{tab:monet}
\end{flushleft}

\vspace{-3mm}
\end{table}

\begin{figure}[t]
\includegraphics[bb=130 49 516 700,height=89mm,angle=-90,clip]{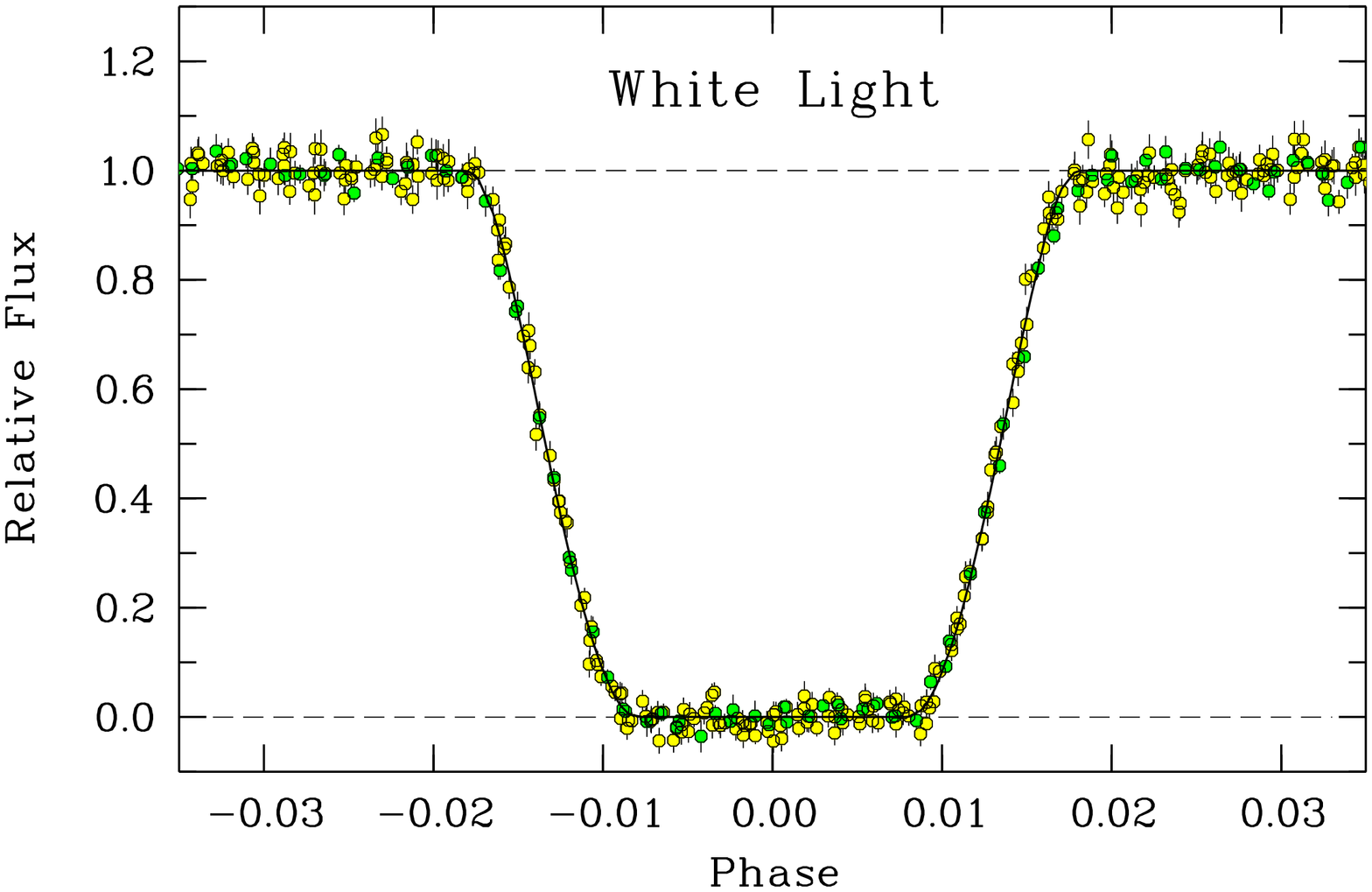}
\includegraphics[bb=130 49 505 700,height=89mm,angle=-90,clip]{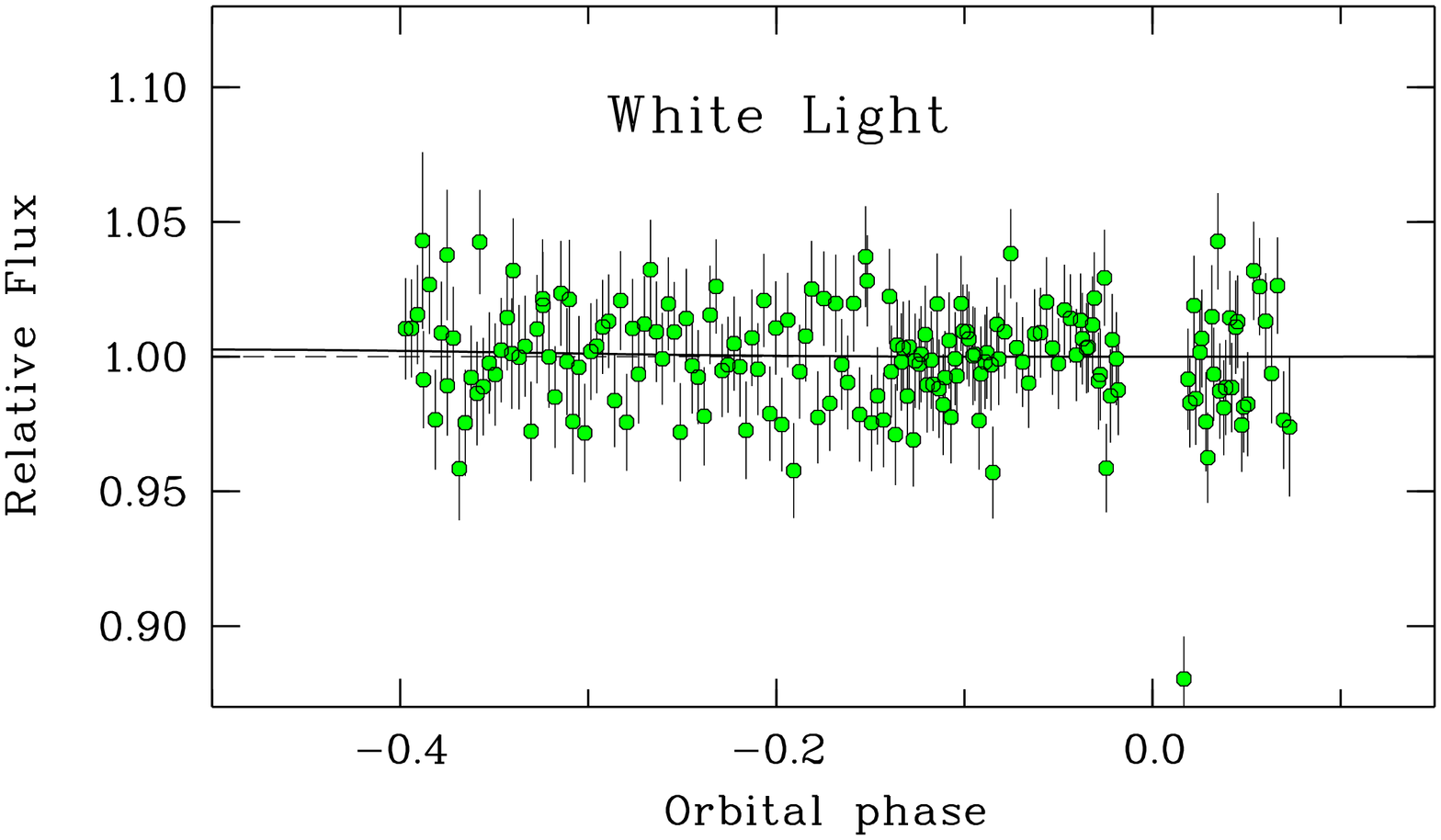}
\includegraphics[bb=130 49 543 700,height=89mm,angle=-90,clip]{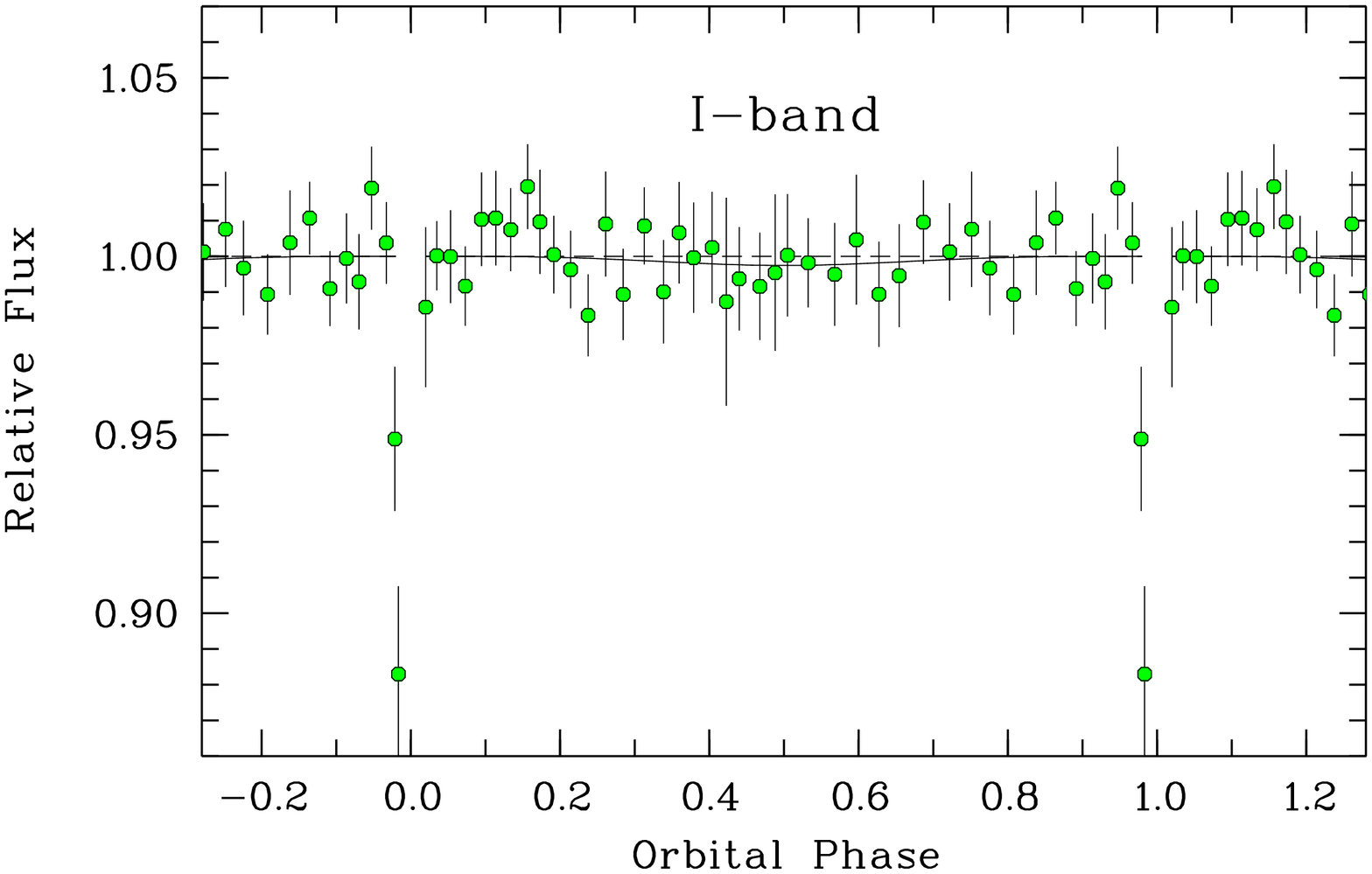}
\caption[chart]{MONET/N white-light and I-band light curves of
  CSS21055 as a function of orbital phase. \emph{Top:} Eclipse light
  curve composed of 10\,s (yellow) and 20\,s (green) exposures. The
  solid line represents the best-fit model light curve for the model
  described in Sect.~\ref{sec:fit}. \emph{Center:} Unbinned orbital
  light curve in WL with 20\,s exposure times. \emph{Bottom:} Binned
  light curve in the Bessell I-band. The folded data are displayed for
  1.6 orbital periods. }
\label{fig:lc}
\end{figure}

\subsection{Folded light curves}
\label{sec:folded}

We folded the white-light and the I-band light curves over the orbital
period of Eq. (1), obtaining phase-resolved light curves. In previous analyses
\citep[e.g.,][]{beuermannetal13}, our model fit allowed for a variation
of the flux outside eclipse. In the present case, no such allowance
was made and we relied on the relative photometry, with the aim to
measure the relative orbital modulation. Before coadding the light
curves, they were normalized to the mean flux in the phase intervals
$\phi\!=\!-0.10$ to $-0.03$ and $\phi\!=\!0.03$ to
$0.10$. Figure~\ref{fig:lc} (top panel) shows the folded eclipse light
curve in WL for the phase interval $\phi\!=\!-0.035$ to 0.035,
composed of twelve light curves obtained under the most favorable
atmospheric conditions. The yellow and green dots represent 10\,s and
20\,s exposures, respectively. The mean relative flux in the
eclipse is $-0.0009\pm0.0025$. The 1-$\sigma$ error corresponds to a
lower limit to the eclipse depth in WL of 6.5 mag. The central panel shows
the coadded WL orbital light curve for cycle
numbers $E\,=\,0$ and 1335. These data cover the phase interval from
$\phi\!=\!-0.397$ to +0.073 with 20\,s exposure times and provide the best
signal-to-noise ratio ($S/N$). The bottom panel shows the phase-folded
orbital light curve in the I-band, which is composed of 908 individual
exposures taken with exposure times between 30 and 180\,s and covering
all orbital phases. We binned the data in 41 phase intervals outside
eclipse and 9 narrower intervals in eclipse. Neither the WL nor the
I-band light curve show evidence for an orbital variation outside the
eclipse (see Sect.~\ref{sec:irrad} for a discussion).

\subsection{I-band flux in eclipse}
\label{sec:ecl}

\begin{figure}[t]
\hspace{17.1mm}\includegraphics[bb=29 172 548 693,height=71.7mm,clip]{22241f2a.ps}

\vspace{-76.7mm} \includegraphics[bb=86 52 546 542,height=89mm,angle=-90,clip]{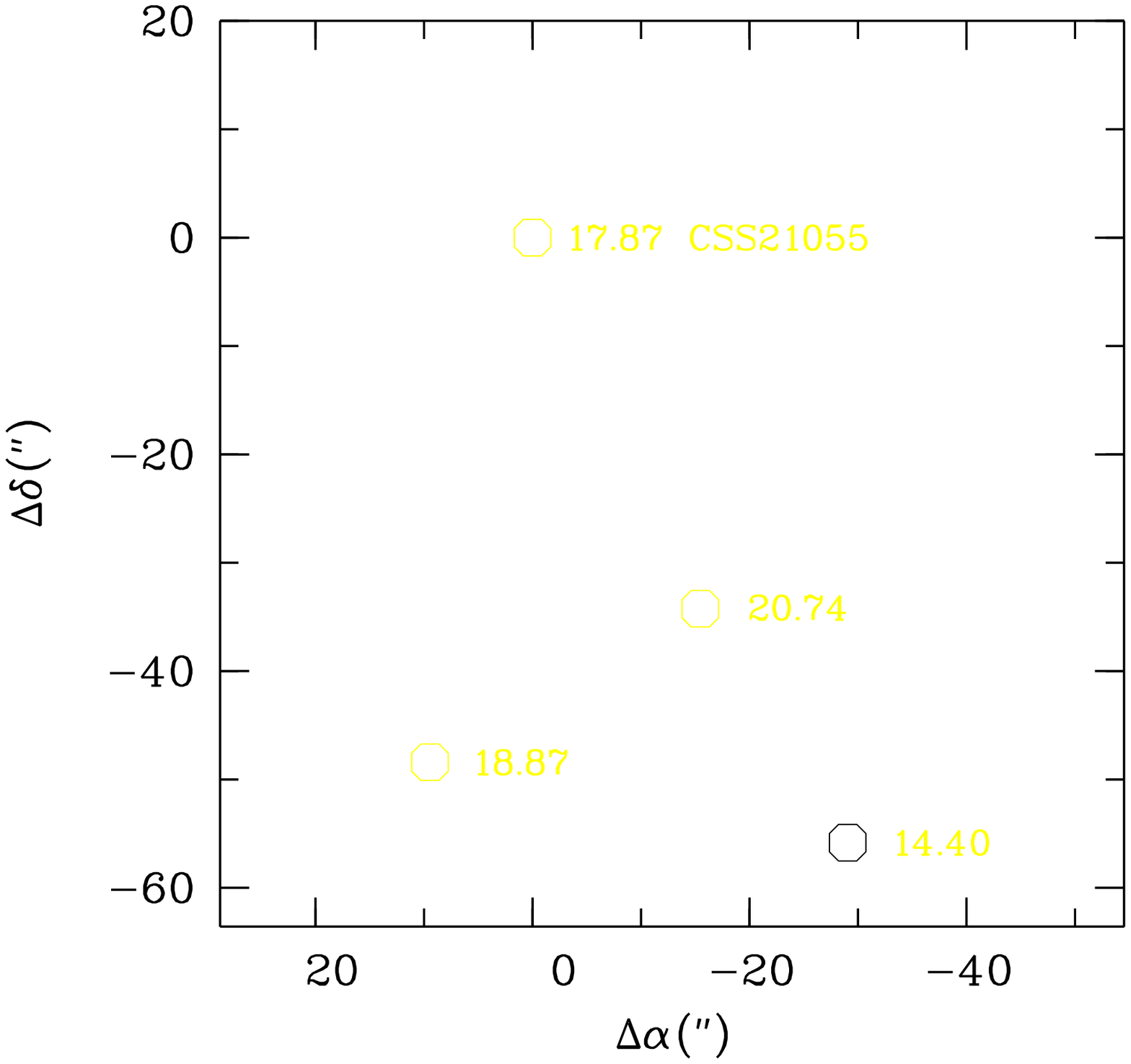}
\caption[chart]{I-band image of the field containing \css\ prior to
  the total eclipse. The image measures $83\arcsec \times 83\arcsec$,
  north is up and east is to the left. The integrated exposure time is
  1200\,s. The quoted I-band magnitudes of stars in the image are
  converted from the Sloan AB magnitudes. The bright star at the
  bottom is comparison star C2. }
\label{fig:out}
\end{figure}

\begin{figure}[t]
\hspace{17.1mm}\includegraphics[bb=29 172 548 693,height=71.7mm,clip]{22241f3a.ps}

\vspace{-76.7mm} \includegraphics[bb=86 52 546 542,height=89mm,angle=-90,clip]{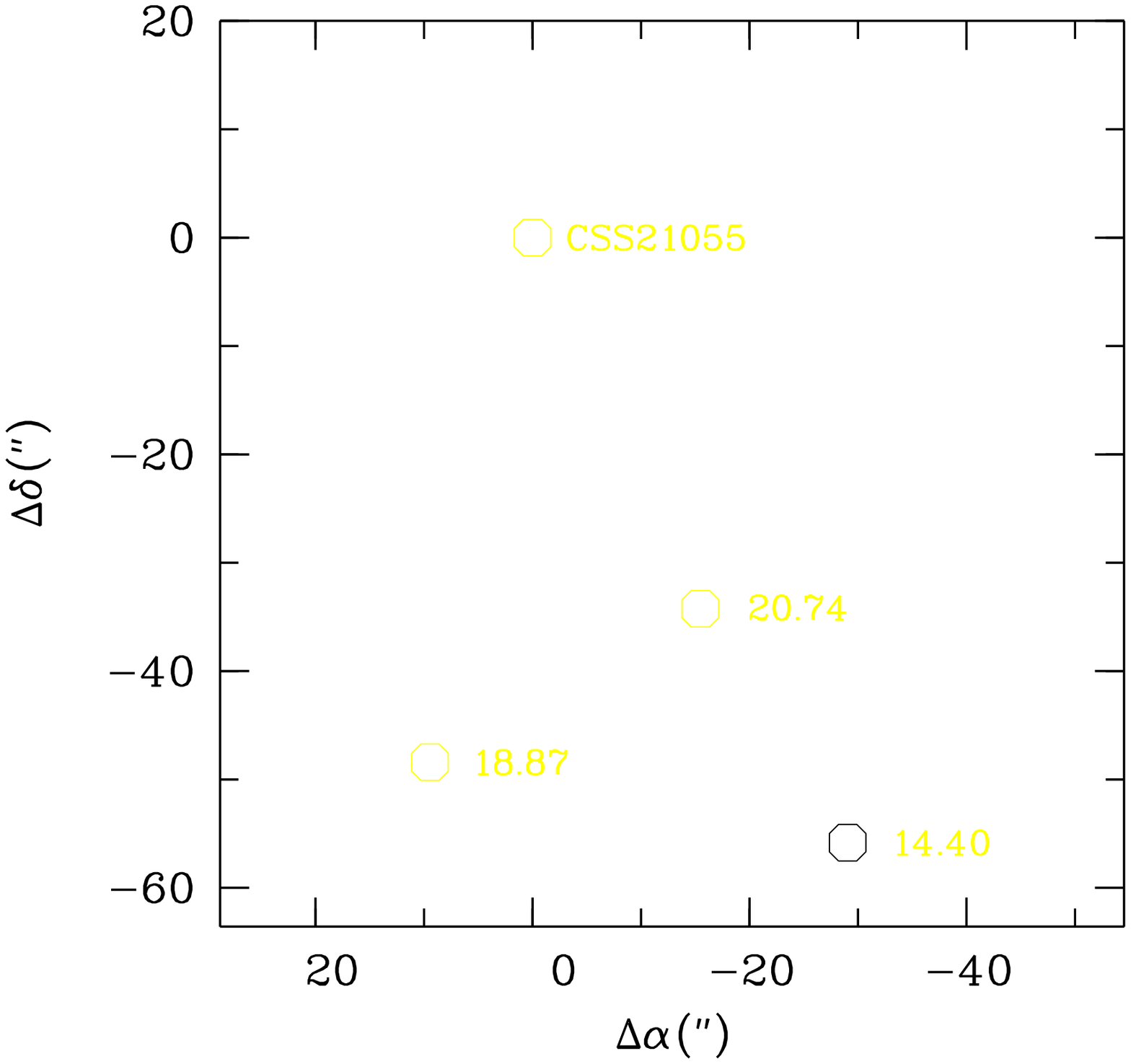}
\caption[chart]{Same as\ Fig.~\ref{fig:out} for an integrated
  exposure time of 600s inside the total eclipse. In both pictures,
  the circles have a radius of one FWHM of the point spread function.}
\label{fig:ecl}
\end{figure}

Instead of averaging the results of the aperture photometry of the
individual exposures inside the eclipse, it is more advantageous to
measure the flux in a coadded image composed of the images taken in
totality. Figures~\ref{fig:out} and \ref{fig:ecl} show the I-band
images accumulated over 1200\,s outside eclipse and 600\,s in the
total eclipse. The eclipse image is composed of sixteen 30\,s
exposures in dark time and two 60\,s exposures taken at full moon, the
out-of-eclipse image contains twice the number of exposures, extracted
from the same five runs (Table~\ref{tab:monet}). Despite the greater
sky brightness, including the longer-exposed moon-time data improves the 
$S/N$ ratio.  The pictures show our 14.4 mag comparison star C2
and two stars fainter than CSS21055. Most of the other sources are
faint galaxies. The I-band magnitudes quoted in the images were
obtained from the SDSS AB magnitudes $r$, $i$, and $z$, using the
transformation equations of Lupton (2005)\footnote{Technical note
  quoted in SDSS DR5, Trans\-for\-ma\-tions between SDSS magnitudes
  and UBVRcIc: http://www.sdss.org/dr5/algorithms/
  sdssUBVRITransform.html}. Based on the SDSS magnitudes quoted above,
CSS21055 has Bessell $I=17.87$ outside eclipse, which calibrates our relative
photometry. The point spread function of the coadded images in
Figs.~\ref{fig:out} and \ref{fig:ecl} has a FWHM of 2.8 pixels. The
faint star below the image center has Sloan $r=21.47$, $i=21.19$, and
$z=21.06$, or $I=20.74$, and is detected in Fig.~\ref{fig:ecl} with a
$S/N=8$ for an extraction radius of 2.8\,pixel. The flux of CSS21055
in eclipse is zero or minimally negative for extraction radii between
2 and 4 pixels. The sky signal in the eclipse image displays a
Gaussian distribution with a standard deviation, which translates to
an error between 0.72\% and 1.00\% of the out-of-eclipse flux of
CSS21055 for extraction radii between 2 and 4 pixels. The 1-$\sigma$
lower limits to the magnitude differences are 5.36 mag and 5.00 mag,
respectively, implying I-band eclipse magnitudes of $I>23.23$ (2
pixel) and $I>22.87$ (4 pixel). Conservatively, we adopt the
latter. The 2-$\sigma$ limit is brighter by 0.75\,mag.

\section{Results}

In this section, we derive the orbital parameters of the binary by
fitting a purely geometrical model to the phase-folded mean eclipse light
curve in WL and and obtain limits on the mass and the absolute
magnitude of the secondary by taking recourse to the WD models of
\citet{holbergbergeron06} and
\citet{tremblayetal11}\footnote{http://www.astro.umontreal.ca/~bergeron/CoolingModels}
and the BD models of \citet{chabrieretal00} and
\citet{baraffeetal03}
\footnote{http://perso.ens-lyon.fr/isabelle.baraffe/COND03$\_$models}.


\subsection{The effective temperature of the white dwarf}
\label{sec:teff}

\begin{figure}[t]
\includegraphics[bb=92 64 554 700,height=89mm,angle=-90,clip]{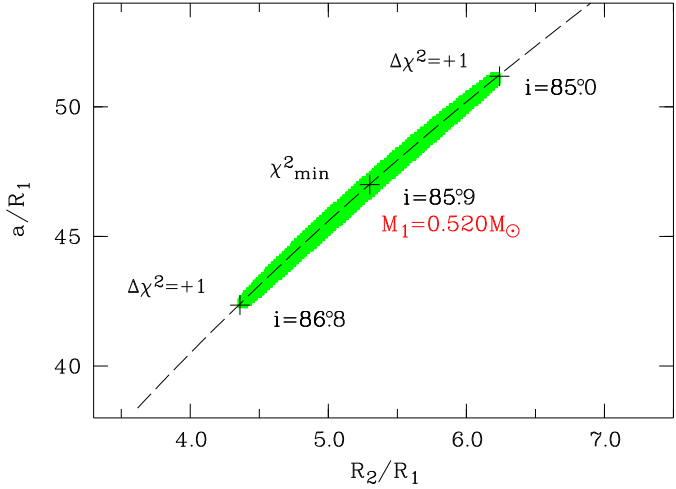}
\caption[chart]{Relation between the best-fit values of $a/R_1$ and
  $R_2/R_1$ for a systematic variation of the inclination $i$ in the
  light-curve fit (dashed curve). The three crosses (+) indicate the
  $\,\chi^2$ minimum and the 1-$\sigma$ error limits. The shaded band
  represents the region with $\Delta\,\chi^2\!<\!1$. The best-fit WD
  mass value is noted in red (see text).}
\label{fig:chi}
\end{figure}

The parameters of the WD derived by \citet{kleinmanetal13} from the
SDSS spectrum and photometry are $T_1\!=\!13074\pm1306$\,K,
log\,$g\!=\!7.89\pm0.12$, and $M_1\!=\!0.551\pm0.058$\,\msun. The
distance is quoted as 176\,pc. Comparison of the WD models with the
observed color $g-i\!=\!-0.39\pm0.01$ translates into a color excess
$E_{g-i}$ between $-0.03$ to 0.14 for the full range of the Kleinman
et al. parameters. The extinction estimated from the Galactic hydrogen
column density $N_\mathrm{H}$ in the direction of \css, suggests a
narrower range of $E_{g-i}$. Using the extinction model of
\citet{cardellietal89} and the calibration between $N_\mathrm{H}$ in
atoms\,cm$^{-2}$ and the visual absorption provided by
\citet{predehlschmitt95}, we find
$E_{g-i}\!=\!0.30\,(N_\mathrm{H}/10^{21})$. For the Galactic
coordinates of CSS21055, $l,b\!=\!16^\circ,+70^\circ$, the NASA
HEASARC
tool\footnote{http://heasarc.gsfc.nasa.gov/cgi-bin/Tools/w3nh/w3nh.pl}
yields a total Galactic column density
$N_\mathrm{H,gal}\!=\!(2.6\pm0.4)\times 10^{20}$\,atoms\,cm$^{-2}$.
The actual column density in front of CSS21055 may be substantially
smaller, but is unlikely to be zero (see Lallement et al. 2003 for the
structure and extent of the local bubble of low-density gas). For
\css\ at $d\!\simeq\!180$\,pc, we estimate that $E_{g-i}$ falls
between 0.03 and 0.09, which implies a WD temperature
$T_1\!\simeq\!13\,000\pm700$\,K for masses $M_1$ between 0.5 and
0.6\,\msun.

\subsection{Geometric-model fit to the mean eclipse light curve}
\label{sec:fit}

We fit the phase-folded WL eclipse light curve (Fig.~\ref{fig:lc}, top
panel) by the geometrical model of a WD eclipsed by the secondary. The
free parameters of the fit are the inclination $i$, the ratio $a/R_1$
of the astrocentric semi-major axis vs. the radius of the primary, the
ratio $R_2/R_1$ of the radii of secondary and primary, and the
surface-brightness ratio $S_2/S_1$ of secondary and primary. The model
is averaged over the finite exposure times of 10\,s or 20\,s before
calculating $\,\chi^2$ and the fit may include a mixed bag of exposure
times. The best-fit model shown in Fig.~\ref{fig:lc} (top panel) is
the one for 10\,s exposures. The model curve for 20\,s exposures
would be slightly more rounded off.

In contrast to Sect.~\ref{sec:ephem}, we now account for limb
darkening of the white dwarf \citep{gianninasetal13}, but keep the
assumption of a uniform disk for the companion, which is black for
practical purposes.  Limb darkening reduces the equivalent
uniform-disk diameter of the WD and the fit compensates this by
increasing the fitted radius $R_1$, with a corresponding reduction in
the implied value of $M_1$ and small changes in all other fit
parameters. We adopted a square-root limb-darkening law
$I(\mu)/I(1)\!=\!1-d(1-\mu)-f(1-\sqrt{\mu})$, where $\mu$ is the
cosine of the angle between the emerging ray and the radial
direction. The parameters $d$ and $f$ were tabulated by
\citet{gianninasetal13} for different photometric bands as functions
of $T_\mathrm{eff}$ and log\,$g$ of the WD. We also tested a quadratic
law, but found no difference in the present fits. As expected,
however, with both laws the fit yields a larger $R_1$ and a smaller
$M_1$ than for the case of no limb darkening.

We estimate that the effective central wavelength for our WL data is
$\lambda_\mathrm{c,WL}\!=\!6000\pm 200$\,\AA, in the blue half of the
Sloan $r$-band. We performed fits with the coefficients $d$ and $f$
for the Sloan bands $g, r$, and $i$, $T_\mathrm{eff}$ between 12000 and
14000\,K, and log\,$g\!=\!7.75$ and 8.00 and found that the combined
systematic error caused by these uncertainties is substantially
smaller than the statistical error. In what follows, we use the
darkening coefficients for the $r$-band,
$T_\mathrm{eff}\!=\!13000$\,K, and log\,$g\!=\!7.75$. The fit yields
$i\!=\!85\fdg9\pm0\fdg9\pm0\fdg1$, $a/R_1\!=\!47.0\pm4.4\pm1.0$, and
$R_2/R_1\!=\!5.30\pm0.94\pm0.18$, where the statistical and the
systematic errors are quoted separately. These parameters are listed
in part $(a)$ of Table~\ref{tab:results} with the two errors added
quadratically. The fit has $\,\chi^2\!=\!752.9$ for 736 degrees of
freedom or a reduced $\,\chi^2_\nu\!=\!1.023$. The best-fit WD mass is
0.52\,\msun\ for the adopted limb-darkening law, compared with
0.58\,\msun\ for no limb darkening. Low $\,\chi^2$ values are obtained
only along a narrow track in the three-dimensional parameter space and
all fit parameters are highly correlated. For $i$ as the independent
variable, $a/R_1$ and $R_2/R_1$ vary as shown by the dashed line in
Fig.~\ref{fig:chi}. The area shaded in green shows where
$\Delta\,\chi^2\!=\!\,\chi^2\!-\!\,\chi^2_\mathrm{min}\!<\!1$ in the
projection of the three-dimensional distribution.

\begin{table}[t]
\begin{flushleft}
\caption{Best-fit parameters for CSS21055 with 1-$\sigma$ correlated
    errors.}
\begin{tabular}{l@{\hspace{9mm}}l@{\hspace{8mm}}l@{\hspace{5mm}}l}
  \hline \hline \\[-1ex]
  Parameter   &  Best Fit         & \multicolumn{2}{l}{~~~~1-$\sigma$ Error} \\[1.0ex]
  \hline\\[-1ex]
  \multicolumn{4}{l}{\emph{(a)~~ Fitted parameters for the geometrical model:}}\\[0.5ex]
  Inclination $i$~~~($^\circ$) & 85.9 & +0.9 & $-0.9$ \\
  $a/R_1$ & 47.0& $-4.5$ & +4.5 \\
  $R_2/R_1$ & 5.3 & $-1.0$ & +1.0\\ [1.3ex]
\multicolumn{4}{l}{\emph{(b)~~ Derived system parameters:}}\\[0.5ex]
$a$~~~($10^{10}$~cm) & 4.70 & $-0.15$ & $+0.15$\\ 
$d$~~~(pc) & 192 & $+9$ & $-9$\\ [1.3ex]
\multicolumn{4}{l}{\emph{(c)~~ Derived parameters of the white dwarf:}}\\[0.5ex]
$T_1$~~~(K) & 13\,000 & $-700$ & $+700$\\
$M_1$~~~(\msun) & 0.520 & $-0.057$ & $+0.057$\\
$R_1$~~~($10^8$~cm) & 10.00 & $+0.69$ & $-0.58$\\ 
log\,$g$~~~ & 7.84 & $-0.10$ & $+0.11$\\
$E_{g-i}$~~~(mag) & 0.06 & $-0.03$ & +0.03 \\
$M_g$~~~(mag) & 11.25 & $-0.04$ & $+0.05$\\ [1.3ex]
\multicolumn{4}{l}{\emph{(d)~~ Derived parameters of the secondary (for $M_2\!=\!0.06$\,\msun): }}\\[0.5ex]
$R_2$~~~($R_\odot$) & 0.076 & $-0.010$ & $+0.009$\\ 
Nightside $I$~~(mag) & $>$22.87&  & $$ \\
Nightside $M_\mathrm{I}$~~(mag) & $>$16.5 & $-0.1$ & $+0.1$\\[1.3ex]
\hline                                                            
\end{tabular}
\label{tab:results}
\end{flushleft}
\end{table}

\subsection{System parameters of \css}
\label{sec:system}

Combining the light-curve fit and the photometric fit yields a set of
internally consistent system parameters. The WD models define $R_1$ as
a function of $M_1$ with a minor dependence on $T_1$. Only $M_1$ and
$R_1$ enter the geometrical fit explicitely via the parameter
$a/R_1$. For simplicity, we assumed that $T_1$ is positively
correlated with $M_1$ on the grounds that an increased mass requires a
higher temperature to maintain the ionization balance in the
line-forming region (see Table~\ref{tab:results}). The semi-major axis
varies as $a\!\propto\!(M_1+M_2)^{1/3}$, which is dominated by the
dependence on $M_1$. The combined fit yields the set of parameters
$a$, $M_1$, $R_1$, $E_{g-i}$, the $g$-band extinction $A_\mathrm{g}$,
the distance $d$, and the absolute magnitude $M_\mathrm{g}$ of the WD,
with errors derived from the uncertainty in $a/R_1$. Sections
$(b)-(d)$ of Table~\ref{tab:results} list the results for
$M_2\!=\!0.06$\,\msun.  As all other parameters, $M_1$ varies
systematically with $i$, assuming values of 0.58\,\msun, 0.52\,\msun,
and 0.46\,\msun\ for inclinations of $85\fdg0$, $85\fdg9$, and
$86\fdg8$, respectively.

\begin{figure}[t]
\includegraphics[bb=91 47 551 721,height=90mm,angle=-90,clip]{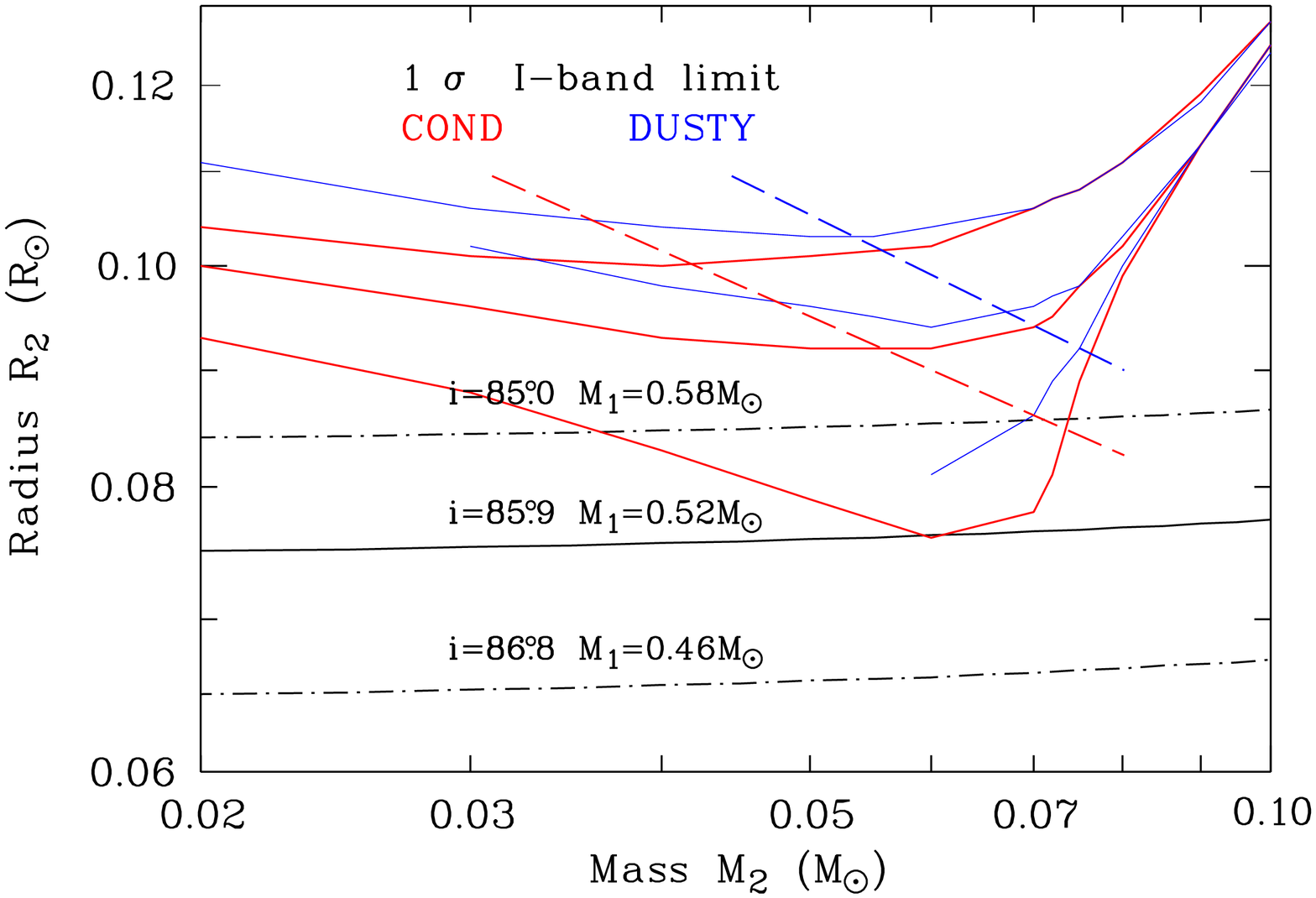}
\caption[chart]{Mass-radius relation of the secondary in \css\
  obtained for the best fit (solid line) and the 1-$\sigma$ limits
  (dot-dashed lines). Also shown are the mass-radius relations for the
  DUSTY models of \citet[][blue curves]{chabrieretal00} and the COND
  models of \citet[][red curves]{baraffeetal03} for ages of 0.5, 1,
  and 10\,Gyr (from top). A secondary with an absolute magnitude
  $M_\mathrm{I}\!>16.5$ is located below the respectively color-coded
  dashed lines.}
\label{fig:R2M2}
\end{figure}

The mass of the secondary is a free parameter that is not effectively
constrained by the fit, but can be limited by the requirement that it
conforms with the (age-dependent) mass-radius relationship $R(M)$ of
low-mass objects.  The temporal evolution of these objects has been
considered by the Lyon group for the two limiting cases of (i) dust
included in the radiative transfer equation (DUSTY models, Chabrier et
al. 2000) and (ii) dust disregarded in the radiative transfer on the
assumption that sedimentation to levels below the photosphere takes
place (COND models, Baraffe et al. 2003). DUSTY models may be more
appropriate for the secondary in \css\ if its temperature exceeds
1300\,K.
In Fig.~\ref{fig:R2M2}, we compare the result of our geometrical model
with the $R(M)$ for the DUSTY and COND models. Our solution overlaps
with the theoretical radii of both models for ages $t\!>\!1$\,Gyr and
$M_2$ between 0.036 and 0.074\,\msun. Our best-fit radius agrees with
that of the COND model for 0.06\,\msun\ and 10\,Gyr, but there is no
model corresponding to the lower masses permitted by our fit. If we
require overlap with the COND or DUSTY model radii, the permitted WD
mass is reduced to 0.52 -- 0.58\,\msun\ at the 1-$\sigma$ level of the
geometrical fit, in excellent agreement with the spectroscopic mass of
$0.551\pm0.058$\,\msun\ \citep{kleinmanetal13}. It is noteworthy that
the lower age limit set by our fit exceeds the cooling age of the WD,
$t_\mathrm{cool}\!=\!0.27\pm0.09$\,Gyr for a temperature of
$13\,000\pm700$\,K and a mass between 0.5 and 0.6\,\msun.

The non-detection of the secondary in the I-band during the eclipse
further restricts $M_2$. The 1-$\sigma$ limit on the absolute
\mbox{I-band} magnitude of the secondary, $M_\mathrm{I}\!>\!$16.5,
defines an age-dependent maximum mass of the DUSTY or COND models
(dashed lines in Fig.~\ref{fig:R2M2}). For an age of 10\,Gyr, the
absolute upper mass limits are 0.074\msun\ and 0.075\,\msun\ for the
COND and DUSTY models, respectively. For an age $t\!=1$\,Gyr, the
corresponding limits are reduced to 0.056\,\msun\ and
0.068\,\msun. The WL flux in eclipse (Sect.~\ref{sec:folded})
corresponds approximately to an absolute R-band magnitude
$M_\mathrm{R}\!\ga18.0$, which supports the above conclusion, but is
less restrictive than the I-band measurement given the steep spectral
slope of the COND and DUSTY models. The lower mass limit is between
0.030 and 0.040\,\msun\ for the COND models, a range where the DUSTY
model may no longer be applicable. Roche lobe overflow is expected
only at still lower masses in the planetary regime. Our results define
the secondary in CSS21055 fairly reliably as a BD with a radius
consistent with the COND or DUSTY models for $t\!>\!1$\,Gyr.

\subsection{Irradiation of the secondary}
\label{sec:irrad}

Irradiation of the secondary by the WD is intense in the substellar
point and subsides as the nightside is entered. 
Most of the incident blue and ultraviolet light, however, will be
Rayleigh scattered from the atmosphere, while the red and infrared
light is mostly absorbed and heats the atmosphere. The details are
complex \cite[e.g.,][]{sudarskyetal03}. The solid angle subtended by
the secondary is $\Omega\!=\!\pi(R_1/a)^2\!\simeq\!0.05$\,sr
(Table~\ref{tab:results}), and with a wavelength-dependent geometrical
albedo $A_\mathrm{g}\!\simeq\!0.6$, 0.3, and $<\!0.05$ at 4000\AA,
6500\AA, and 8000\AA\ \citep{gelinoetal99}, we expect a relative
orbital modulation of the reflected light
$f_\mathrm{sec,refl}(\lambda)/f_\mathrm{wd}(\lambda)\!=\!A_\mathrm{g}\Omega/\pi$
of 0.9\%, 0.4\%, and $<\!0.1$\% in the blue, red, and infrared
bands.

Reprocessing of absorbed radiation leads to a position-dependent rise
in the effective temperature. A rough estimate is obtained from
the average equilibrium temperature $T_\mathrm{2,eq}$ in the heated
hemisphere,
\begin{equation}
T_\mathrm{2,eq}^4 \simeq T_2^4+T_1^4(R_1^2/2a^2)(1-A_\mathrm{B}),
\end{equation}
where $A_\mathrm{B}$ is the Bond albedo and the factor of 1/2 in the
bracket accounts for the difference between the cross section and the
emitting area of the irradiated hemisphere. For the best-fit
parameters in Table~\ref{tab:results} and a Bond albedo of 0.5
\citep{marleyetal99}, the effective temperature rises moderately from
the nightside value $T_2\!\la\,$1800\,K for the adopted limit of 1\%
of the WD flux in the I-band to an average 1925\,K in the irradiated
hemisphere. From the COND models of \citet{baraffeetal03}, we find
that the surface brightness in the I-band increases by 1.0\,mag for a
temperature rise of 250\,K, implying a meagre difference between the
dark and irradiated hemispheres of only about 0.5\,mag and an expected
orbital modulation of about 0.55\% relative to the I-band flux of the
WD. A similar result is obtained for the DUSTY models.
Not surprisingly, our observed light curves in WL and in the I-band do
not show a measurable enhancement around orbital phase $\phi\!=\!0.5$ due to
either effect. We have fitted a light curve calculated with the
irradiation model of \citet{beuermannreinsch08} to the data and
obtained an amplitude of $0.26\pm0.11\%$ for the (unfortunately
incomplete) WL light curve and of $-0.25\pm0.42$\% for the I-band
(solid black curves in Fig~\ref{fig:lc}, central and bottom
panels). Both results are consistent with the BD nature of the
secondary as suggested by the non-detection in the I-band eclipse.

\section{Discussion}
\label{sec:disc}

We have found that CSS21055 is a totally eclipsing short-period
post-common envelope binary with an orbital period of 121.73\,min,
containing a WD primary and a probable BD secondary. The likely mass
of the companion is between 0.030 and 0.074\,\msun. A planet-sized
body is not supported by the fits and would, moreover, overfill its
Roche lobe. WD+BD binaries are rare \citep{steeleetal11} and only two
such systems with comparable short orbital periods are known,
WD0137+349B with 116\,min period \citep{maxtedetal06,burleighetal06}
and NLTT5306 with 101.88\,min \citep{steeleetal13}, and both are not
eclipsing. Other WD+BD binaries exist in the variety of cataclysmic
variables, but their post-CE evolutionary history was probably
different.

The BD nature of the secondary in CSS21055 is suggested by its
non-detection in the Bessell I-band during the total eclipse of the
WD. At a distance of $\sim\!190$\,pc, the 1-$\sigma$ limit on the
absolute magnitude $M_\mathrm{I}\!\ga\!16.5$ implies a nightside
temperature $T_2\!\la\!1800$\,K based on the COND models of
\citep{baraffeetal03} or $T_2\!\la\!1980$\,K for the DUSTY models of
\citet{chabrieretal00}. For a BD mass of 0.06\,\msun, the
radial-velocity amplitude of the WD would be about
40\,km\,s$^{-1}$. It is possible that the broad absorption lines of
the WD contain narrow emission-line cores from the heated face of the
secondary. The single available SDSS spectrum is not conclusive in
this respect. Our accurate ephemeris will permit to probe the
nightside atmosphere of the companion spectroscopically in the 125\,s
of the total eclipse of the WD, using an 8-m class telescope. If the
companion is near the upper mass limit $M_2\!\simeq\!0.073$\,\msun,
the colors of the COND models suggest that it could reach 18.9\,mag in
$K$ and about 17.5\,mag in the warm Spitzer 3.6$\,\mu$ and 4.5$\,\mu$
bands. Even at 0.04\,\msun, the apparent brightness of $\sim\!20$\,mag
may still allow detection with Spitzer.

A comparison of the observed brightness limit with the predictions of the
COND and DUSTY models suggests that the secondary has an age exceeding
1\,Gyr, which is larger than the cooling age of the WD of
$0.27\pm0.09$\,Gyr for a temperature of $13\,000\pm700$\,K and a mass
between 0.5 and 0.6\,\msun. If true, the companion to the WD existed
before the CE event. The spectroscopically and geometrically deduced
mass argues for a carbon-oxygen WD born on the asymptotic giant
branch. Modeling the dispersal of the envelope and the formation of
the close binary may present a challenge to CE theory
\citep[e.g.,][]{diehletal08}.

\begin{acknowledgements}
  Most of our data were obtained with the MONET/North telescope of the
  MOnitoring NEtwork of Telescopes, funded by the Alfried Krupp von
  Bohlen und Halbach Foundation, Essen, and operated by the
  Georg-August-Universit\"at G\"ottingen, the McDonald Observatory of
  the University of Texas at Austin, and the South African
  Astronomical Observatory.  The ``Astronomie \& Internet'' program of
  the Foundation and the MONET consortium provide observation time to
  astronomical projects in high schools worldwide. In addition, this
  paper uses observations made at the South African Astronomical
  Observatory (SAAO). We thank the Calar Alto Observatory for the
  allocation of director's discretionary time to this programme.
  Funding for SDSS-III has been provided by the Alfred P. Sloan
  Foundation, the Participating Institutions, the National Science
  Foundation, and the U.S. Department of Energy Office of Science. The
  SDSS-III web site is http://www.sdss3.org/.  SDSS-III is managed by
  the Astrophysical Research Consortium for the Participating
  Institutions of the SDSS-III worldwide Collaboration.
\end{acknowledgements}

\bibliographystyle{aa}

\end{document}